\documentclass[10pt]{article}
\usepackage[cp866]{inputenc}
\begin{document}

\centerline {{\Large\bf Role of exterior and evolutionary skew-symmetric }}
\centerline {{\Large\bf differential forms in mathematical physics.}}
\centerline {L.I. Petrova}
\centerline{{\it Moscow State University, Russia, e-mail: ptr@cs.msu.su}}
\bigskip

A role of skew-symmetric differential forms in mathematical physics
relates to the fact that they reflect the properties of conservation
laws. The closed exterior forms correspond to the conservation laws for
physical fields, whereas the evolutionary forms correspond 
to the conservation laws for material systems - material media. (The 
conservation laws for physical fields are the exact conservation 
laws - conjugated objects, whereas the conservation laws for material 
media are the balance, differential, conservation laws.)

Skew-symmetric differential forms possess an unique properties, namely, 
they can describe a conjugacy of any objects (for example, derivatives 
with respect to various variables, differential equations 
composing a set of equations, and so on). 
Such a possibility is due to that the skew-symmetric differential forms, 
as opposed to differential equations, deal with differentials 
and differential equations rather than with derivatives. 
The closed exterior forms describe conjugated objects. It is precisely 
these conjugated objects that correspond to the exact conservation laws.  
And the evolutionary forms, whose basis (in contrast to the exterior 
forms) are deforming manifolds, describe 
the process of conjugating objects and obtaining conjugated 
objects - the closed exterior forms. 
That is, the evolutionary forms discribe the process of obtaining closed 
exterior forms.  

It is just with this peculiarity the role of evolutionary forms in 
mathematical physics is connected.

Since from the evolutionary forms, which correspond to the  
conservation laws for material system, one obtains the closed
exterior forms, which correspond to the conservation laws
for physical fields, hence it follows that  physical fields are
generated by material systems. 

The relation between evolutionary forms and closed exterior ones 
discloses the relation between the equations of mathematical physics 
describing evolutionary processes in material media and field theories 
describing physical fields. This explains the field theory postulates.

Transition from evolutionary forms, which are unclosed ones,
to the closed exterior forms is possible only under degenerate
transformation, conditions of which are symmetries of dual forms. 
Under describing physical processes in material media such symmetries 
are due to the degrees of freedom of material media. 

Symmetries of dual forms are external symmetries of the equations of 
field theories. And symmetries of closed exterior forms, which are  
conditions of fulfilment of the exact conservation laws, are 
interior symmetries of field theories.  From this one can see a 
connection between internal and external symmetries.

\bigskip
{\large\bf Distinction between exterior and evolutionary forms}

The distinction between exterior and evolutionary skew-symmetric
differential forms is connected with the properties of manifolds
on which skew-symmetric forms are defined.

It is known that the exterior differential forms [1] are skew-symmetric
differential forms whose basis are differentiable manifolds or they can be
manifolds with structures of any type. Structures, on which exterior forms
are defined, have closed metric forms.

It has been named as evolutionary forms the skew-symmetrical differential forms
whose basis are deforming manifolds, i.e. manifolds with
unclosed metric forms. The metric form differential, and correspondingly
its commutator, are nonzero. Commutators of metric forms of such
manifolds describe the manifold deformation (torsion, curvature and so on).

Lagrangian manifolds, manifolds constructed of trajectories of material
system elements, tangent manifolds of differential equations describing
physical processes and others can be examples of deforming manifolds.

A specific feature of evolutionary forms, i.e skew-symmetric forms
defined on deforming manifolds, is the fact that commutators of these forms
include commutators of the manifold metric forms being nonzero. Such commutators
possess the evolutionary and topological properties. Just due to such properties
of commutators  evolutionary forms can generate closed inexact exterior 
forms. 

It is known that the exterior differential form of degree $p$ ($p$-form)
can be written as [2,3]
$$
\theta^p=\sum_{i_1\dots i_p}a_{i_1\dots i_p}dx^{i_1}\wedge
dx^{i_2}\wedge\dots \wedge dx^{i_p}\quad 0\leq p\leq n\eqno(1)
$$
Here $a_{i_1\dots i_p}$ are functions of the variables $x^{i_1}$,
$x^{i_2}$, \dots, $x^{i_p}$, $n$ is the dimension of space,
$\wedge$ is the operator of exterior multiplication, $dx^i$,
$dx^{i}\wedge dx^{j}$, $dx^{i}\wedge dx^{j}\wedge dx^{k}$, \dots\
is the local basis which satisfies the condition of exterior
multiplication.

The differential of the (exterior) form $\theta^p$ is expressed as
$$
d\theta^p=\sum_{i_1\dots i_p}da_{i_1\dots
i_p}\wedge dx^{i_1}\wedge dx^{i_2}\wedge \dots \wedge dx^{i_p} \eqno(2)
$$
The evolutionary differential form of degree $p$ ($p$-form) is written
similarly to exterior differential form. But the evolutionary form
differential cannot be written similarly to that presented for
exterior differential forms. In the evolutionary form
differential there appears an additional term connected with the fact
that the basis of evolutionary form changes. For differential
forms defined on the manifold with unclosed metric form one has
$d(dx^{\alpha_1}\wedge dx^{\alpha_2}\wedge \dots \wedge dx^{\alpha_p})\neq 0$.
(For differential forms defined on the manifold with closed metric form
one has $d(dx^{\alpha_1}\wedge dx^{\alpha_2}\wedge \dots \wedge dx^{\alpha_p})=0$).
For this reason the differential of the evolutionary form $\theta^p$ can be
written as
$$
d\theta^p{=}\!\sum_{\alpha_1\dots\alpha_p}\!da_{\alpha_1\dots\alpha_p}\wedge
dx^{\alpha_1}\wedge dx^{\alpha_2}\dots \wedge
dx^{\alpha_p}{+}\!\sum_{\alpha_1\dots\alpha_p}\!a_{\alpha_1\dots\alpha_p}
d(dx^{\alpha_1}\wedge dx^{\alpha_2}\dots \wedge dx^{\alpha_p})\eqno(3)
$$
where the second term is a differential of unclosed metric form being
nonzero.

[In further presentation the symbol of summing $\sum$ and the symbol
of exterior multiplication $\wedge$ will be omitted. Summation
over repeated indices will be implied.]

The second term connected with the differential of the basis can be 
expressed in terms of the metric form commutator.

For example, let us consider the first-degree form
$\theta=a_\alpha dx^\alpha$. The differential of this form can
be written as
$$d\theta=K_{\alpha\beta}dx^\alpha dx^\beta\eqno(4)$$
where
$K_{\alpha\beta}=a_{\beta;\alpha}-a_{\alpha;\beta}$ are the
components of commutator of the form $\theta$, and
$a_{\beta;\alpha}$, $a_{\alpha;\beta}$ are covariant
derivatives. If we express the covariant derivatives in terms of
the connectedness (if it is possible), they can be written
as $a_{\beta;\alpha}=\partial a_\beta/\partial
x^\alpha+\Gamma^\sigma_{\beta\alpha}a_\sigma$, where the first
term results from differentiating the form coefficients, and the
second term results from differentiating the basis. We arrive at the
following expression for the commutator components of the form $\theta$
$$
K_{\alpha\beta}=\left(\frac{\partial a_\beta}{\partial
x^\alpha}-\frac{\partial a_\alpha}{\partial
x^\beta}\right)+(\Gamma^\sigma_{\beta\alpha}-
\Gamma^\sigma_{\alpha\beta})a_\sigma\eqno(5)
$$
Here the expressions
$(\Gamma^\sigma_{\beta\alpha}-\Gamma^\sigma_{\alpha\beta})$
entered into the second term are just the components of
commutator of the first-degree metric form.

If to substitute the expressions (5) for evolutionary form
commutator into formula (4), we obtain the
following expression for differential of the first degree
skew-symmetric form
$$
d\theta=\left(\frac{\partial a_\beta}{\partial
x^\alpha}-\frac{\partial a_\alpha}{\partial
x^\beta}\right)dx^\alpha dx^\beta+\left((\Gamma^\sigma_{\beta\alpha}-
\Gamma^\sigma_{\alpha\beta})a_\sigma\right)dx^\alpha dx^\beta\eqno(6)
$$
The second term in the expression for differential of skew-symmetric 
form is connected with the differential of the manifold metric form, 
which is expressed in terms of the metric form commutator.

Thus, the differentials and, correspondingly, the commutators of
exterior and evolutionary forms are of different types. In
contrast to the exterior form commutator, the evolutionary form
commutator includes two terms. These two terms have different
nature, namely, one term is connected with coefficients of the
evolutionary form itself, and the other term is connected with
differential characteristics of manifold. Interaction between
terms of the evolutionary form commutator (interactions between
coefficients of evolutionary form and its basis) provides the
foundation of evolutionary processes that lead to generation of
closed inexact exterior forms to which the differential-geometrical
structures are assigned.

\bigskip
{\large\bf Closed inexact exterior forms. Conservation laws}

From the closure condition of the exterior form $\theta^p$:
$$
d\theta^p=0\eqno(7)
$$
one can see that the closed exterior form $\theta^p$ is a conserved
quantity. This means that this can correspond to a conservation law,
namely, to some conservative quantity.

If the form is closed only on pseudostructure, i.e. this form is 
a closed inexact one, the closure conditions are written as
$$
d_\pi\theta^p=0\eqno(8)
$$
$$
d_\pi{}^*\theta^p=0\eqno(9)
$$
where ${}^*\theta^p$ is the dual form.

Condition (9), i.e. the closure condition for dual form, specifies
the pseudostructure $\pi$.
\{Cohomology (de Rham cohomology,
singular cohomology), sections of cotangent bundles and
so on may be regarded as examples of pseudostructures.\}
From conditions (8) and (9) one can see the following. The dual form
(pseudostructure) and closed inexact form (conservative quantity)
made up a conjugated, conservative object that can also correspond to 
some conservation law.
The conservative object, which corresponds to the conservation law,
is a differential-geometrical structure. Such
differential-geometrical structures are examples of G-structures.
The physical structures, which forms physical
fields, and corresponding conservation laws are just such structures.

\bigskip
{\bf Properties of closed exterior differential forms.}

1. {\it Invariance.}

It is known that the closed exact form is
a differential of the form of lower degree:
$$
\theta^p=d\theta^{p-1}\eqno(10)
$$
Closed inexact form is also a differential, and yet not a total one but
an interior on pseudostructure
$$
\theta^p_\pi=d_\pi\theta^{p-1}\eqno(11)
$$

Since closed exterior differential forms are differentials,
they turn out to be invariant under all transformations that conserve
the differential. Gauge transformations (the unitary, tangent, 
canonical, gradient, and other transformations) are examples
of such nondegenerate transformations under which closed exterior forms 
turn out to be invariant.

2. {\it Conjugacy. Duality. Symmetries.}

The closure of exterior differential forms is the result of conjugacy of
elements of exterior or dual forms. The closure property of the exterior
form means that any objects, namely, elements of the exterior form,
components of elements, elements of the form differential, exterior and
dual forms and others, turn out to be conjugated.

With the conjugacy it is connected the duality.

Examples: the conjugacy of coefficients of the form, that of coefficients of
the differential; the conjugacy and duality of the forms of sequential
degrees or of the exterior and dual forms. (Duality:  the closed exterior form
is a conservative quantity and the closed form can correspond to a certain
potential force.)

The conjugacy is possible if there is one or another type of symmetry.

The gauge symmetries, which are interior symmetries of field theory and
with which gauge transformations are connected, are symmetries
of closed exterior differential forms.

The conservation laws for physical fields are connected with such
interior symmetries.

\bigskip
{\bf Identical relations of exterior forms.}

Since the conjugacy is a certain connection between two operators or
mathematical objects, it is evident that, to express a conjugacy
mathematically, it can be used relations.
These are identical relations.

The identical relations express the fact that each closed exterior
form is the differential of some exterior form (with the degree less
by one). In general form such an identical relation can be written as
$$
d _{\pi}\varphi=\theta _{\pi}^p\eqno(12)
$$

In this relation the form in the right-hand side has to be a
{\it closed} one.

Identical relations of exterior differential forms are a mathematical
expression of various types of conjugacy that leads to closed exterior
forms.

Such relations like the Poincare invariant, vector and tensor identical
relations, the Cauchi-Riemann conditions, canonical relations, the integral
relations by Stokes or Gauss-Ostrogradskii, the thermodynamic
relations, the eikonal relations, and so on are examples of identical relations
of closed exterior forms that have the form of relation (12) or its differential
or integral analogs.

One can see that identical relations of closed exterior differential forms
make itself evident in various branches of physics and mathematics.

Below the mathematical and physical meaning of these relations and their role
in generating physical structures will
be disclosed with the help of evolutionary forms.

\bigskip
{\large\bf Role of exterior forms in field theory.}

The properties of closed exterior differential forms correspond to the
conservation laws for physical fields. Therefore, the mathematical
principles of the theory of closed exterior differential forms lie at
the basis of existing field theories.

The properties of closed exterior forms, namely, invariance, conjugacy and duality,
lie at the basis of the group and structural properties of field theory.

The nondegenerate transformations of field theory are transformations of
the closed exterior forms. As it has been pointed out, the gauge
transformations like the unitary, tangent, canonical, gradient and other
transformations are such transformations. There are transformations conserving
the differential. Applications of nondegenerate transformations to identical
relations enables one to find new closed exterior forms and hence to find
new physical structures.

The gauge, i.e. internal, symmetries of the field theory equations are those
of closed exterior forms.

The nondegenerate transformations of exterior differential forms lie at the basis
of field theory operators.
If, in addition to the exterior differential, we
introduce the following operators: (1) $\delta$ for transformations that convert
the form of $(p+1)$ degree into the form of $p$ degree, (2) $\delta'$
for cotangent transformations, (3) $\Delta$ for the
$d\delta-\delta d$ transformation, (4)$\Delta'$ for the $d\delta'-\delta'd$
transformation, one can write down the operators in the field
theory equations in terms of these operators that act on the exterior
differential forms. The operator $\delta$ corresponds to Green's
operator, $\delta'$ does to the canonical transformation operator,
$\Delta$ does to the d'Alembert operator in 4-dimensional space, and
$\Delta'$ corresponds to the Laplace operator. One can see
that the operators of the exterior differential form theory are
connected with many operators of mathematical physics.

It can be shown that the equations of existing field theories are
those obtained on the basis of the properties of the exterior form
theory. The Hamilton formalism is based on the properties of closed
exterior and dual forms of the first degree, quantum mechanics does on
the forms of zero degree, the electromagnetic field equations are based
on the forms of second degree. The third degree forms are assigned to
the gravitational field.

The connection between the field theory equations and closed
exterior forms shows that to every physical field it is assigned
the appropriate degree of closed exterior form. The type of gauge
transformations used in field theory depends on the degree of
closed exterior differential form.
This shows that it is possible to introduce a classification of physical
fields according to the degree of closed exterior form.
(But within the framework of only exterior differential forms one cannot
understand how this classification is explained. This can be elucidated
only by application of evolutionary differential forms.)

\bigskip
Thus one can see that the properties and mathematical apparatus of closed exterior
forms made up the basis of existing field theories. (It should be emphasized
that the field theories are connected with the properties of
{\it inexact} closed exterior forms.)

It is known that the equations of existing field theories have been obtained
on the basis of postulates. One can see that these postulates are obtained from
the closure conditions of inexact exterior forms.

And here it arises the question of  how closed inexact exterior
forms, which correspond to physical structures and
reflect the properties of conservation laws, are obtained. This gives the
answers to the following questions: (a) how the physical structures originate;
(b) what generates physical structures; (c) how the process of generation
proceeds, and (d) what is responsible for
such processes? That is, this enables one to understand the heart of physical
evolutionary processes and their causality.

Below it will be shown that the closed inexact exterior forms can be obtained
from the evolutionary forms.

\bigskip
{\large\bf Properties  of evolutionary forms.}

Above it has been shown that the evolutionary form commutator
includes the commutator of the manifold metric form which is nonzero.
Therefore, the evolutionary form commutator cannot be equal to zero.
This means that the evolutionary form
differential is nonzero. Hence, the evolutionary form, in
contrast to the case of the exterior form, cannot be closed. This leads to
that in the mathematical apparatus of evolutionary forms there arise
new unconventional elements like nonidentical relations and degenerate
transformations. Just such peculiarities allow to describe
evolutionary processes.

Nonidentical relations of the evolutionary form theory, as well as
identical relations of the theory of closed exterior forms,
are relations between the differential and the skew-symmetric form.
In the right-hand side of the identical relation of exterior forms
(see relation (12))  it stands a closed form, which is
a differential as well as the left-hand side. And in the right-hand of the
nonidentical relation of evolutionary form it stands
the evolutionary form that is not closed and, hence, cannot be a differential
like the left-hand side.
Such a relation cannot be identical one.

Nonidentical relations are obtained while describing any processes.
The relation of such type is obtained, for example, while analyzing the integrability
of the partial differential equation. An equation is integrable
if it can be reduced to the form $d\phi=dU$. However it
appears that, if the equation is not subject to an additional
condition (the integrability condition), the right-hand side turns out to be
an unclosed form and it cannot be expressed as a differential.

Nonidentical relations of evolutionary forms are evolutionary relations
because they include the evolutionary form.
Such nonidentical evolutionary relations appear to be selfvarying
ones.  A variation of any object of the relation in some
process leads to a variation of another object and, in turn, a variation
of the latter leads to a variation of the former. Since one of the objects
is a noninvariant (i.e. unmeasurable) quantity, the other cannot be compared
with the first one, and hence,
the process of mutual variation cannot be completed.

The nonidentity of evolutionary relation is connected with a
nonclosure of the evolutionary form, that is, it is connected with
the fact that the evolutionary form commutator is nonzero.
As it has been pointed out, the evolutionary form commutator includes two terms:
one term specifies the mutual variations of the evolutionary form
coefficients, and the second term (the metric form commutator) specifies
the manifold deformation. These terms have a different nature and cannot
make the commutator vanish. In the process of selfvariation of the nonidentical
evolutionary relation it proceeds an exchange between the terms of
the evolutionary form commutator and this is realized according to the
evolutionary relation. The evolutionary
form commutator describes a quantity that is a moving force of the evolutionary
process and leads to generation of differential-geometrical structures.

The significance of the evolutionary relation selfvariation consists in
the fact that in such a process it can be realized the conditions of degenerate
transformation under which the closed inexact exterior form is obtained from
the evolutionary form, and from nonidentical relation the identical relation
is obtained.

\bigskip
Thus, one can see that the closed exterior forms, which possess invariant
properties, are obtained from the evolutionary forms, i.e. skew-symmetric
forms defined on deforming manifolds.

Below it will be shown a role of evolutionary forms in mathematical physics,
which describes physical processes in material media. With the help of
evolutionary forms it will be shown how in material media the physical structures,
which made up physical fields, emerge. This will disclose a role of evolutionary forms
in field theory.

[Sometimes below it will be used  double notation in subtitles, one in reference
to physical meaning and another in reference to mathematical meaning.]

\bigskip
{\large\bf Role of evolutionary forms in mathematical physics and field
theory}

The role of evolutionary forms in mathematical physics and field theory
(as well as the role of exterior forms) is due to the fact that they
correspond to conservation laws. However, these conservation laws are those
not for physical fields but for material media. These are balance
conservation laws for material media,i.e. for material systems.
\{Material system is a variety of elements that have internal structure
and interact to one another. Thermodynamic and
gas dynamical systems, systems of charged particles, cosmic systems, systems
of elementary particles and others are examples of material systems.\}

\bigskip
{\bf Connection of evolutionary forms with the balance conservation laws.}

The balance conservation laws are conservation laws that
establish a balance between the variation of physical quantity and
the corresponding external action. They are described by differential equations.
The balance conservation laws for material systems are conservation laws
for energy, linear momentum, angular momentum, and
mass. From the equations, which describe the balance conservation laws,
one can obtain a relation that is nonidentical relation since it contains
the evolutionary forms.

Let us analyze the equations that describe the balance conservation
laws for energy and linear momentum.

In the accompanying reference system (this system is connected with the manifold
built by the trajectories of the material system elements) the energy
equation is written in the form
$$
{{\partial \psi }\over {\partial \xi ^1}}\,=\,A_1 \eqno(13)
$$

Here $\psi$  is the functional specifying the state of material system
(the action functional, entropy or wave function can be regarded as
examples of such a functional), $\xi^1$ is the coordinate along the
trajectory, $A_1$ is the quantity that depends on specific features of
material system and on external energy actions onto the system.

In a similar manner, in the accompanying reference system the
equation for linear momentum appears to be reduced to the equation of
the form
$$
{{\partial \psi}\over {\partial \xi^{\nu }}}\,=\,A_{\nu },\quad \nu \,=\,2,\,...\eqno(14)
$$
where $\xi ^{\nu }$ are the coordinates in the direction normal to the
trajectory, $A_{\nu }$ are the quantities that depend on the specific
features of material system and on external force actions.

Eqs. (13) and (14) can be convoluted into the relation
$$
d\psi\,=\,A_{\mu }\,d\xi ^{\mu },\quad (\mu\,=\,1,\,\nu )\eqno(15)
$$
where $d\psi $ is the differential
expression $d\psi\,=\,(\partial \psi /\partial \xi ^{\mu })d\xi ^{\mu }$.

Relation (15) can be written as
$$
d\psi \,=\,\omega \eqno(16)
$$
here $\omega \,=\,A_{\mu }\,d\xi ^{\mu }$ is the skew-symmetrical differential form of the first degree.

Since the balance conservation laws are evolutionary ones, the relation
obtained is also an evolutionary relation.

Relation (16) was obtained from the equation of the balance conservation
laws for energy and linear momentum. In this relation the form $\omega $
is that of the first degree. If the equations of the balance conservation
laws for angular momentum be added to the equations for energy and linear
momentum, this form in the evolutionary relation will be a form of the
second degree. And in  combination with the equation of the balance
conservation law for mass this form will be a form of degree 3.

Thus, in general case the evolutionary relation can be written as
$$
d\psi \,=\,\omega^p \eqno(17)
$$
where the form degree  $p$ takes the values $p\,=\,0,1,2,3$..
(The evolutionary relation for $p\,=\,0$ is similar to that in the differential
forms, and it was obtained from the interaction of energy and time.)

Let us show that relation obtained from the equation
of the balance conservation laws proves to be nonidentical.

To do so we shall analyze relation (16).

In the left-hand side of relation (16) there is a
differential that is a closed form. This form is an invariant
object. The right-hand side of relation (16) involves the differential
form $\omega$ that is not an invariant object because in real processes,
as it will be shown below, this form proves to be unclosed. The commutator of this
form is nonzero. The components of commutator of the form $\omega \,=\,A_{\mu }d\xi ^{\mu }$
can be written as follows:
$$
K_{\alpha \beta }\,=\,\left ({{\partial A_{\beta }}\over {\partial \xi ^{\alpha }}}\,-\,
{{\partial A_{\alpha }}\over {\partial \xi ^{\beta }}}\right )
$$
(here the term connected with the manifold metric form
has not yet been taken into account).

The coefficients $A_{\mu }$ of the form $\omega $ have been obtained either
from the equation of the balance conservation law for energy or from that for
linear momentum. This means that in the first case the coefficients depend
on the energetic action and in the second case they depend on the force action.
In actual processes energetic and force actions have different nature and appear
to be inconsistent. The commutator of the form $\omega $ consisted of
the derivatives of such coefficients is nonzero.
This means that the differential of the form $\omega $
is nonzero as well. Thus, the form $\omega$ proves to be unclosed and
cannot be a differential like the left-hand side.
This means that relation (16) cannot be an identical one.

In a similar manner one can prove the nonidentity of relation (17).

\bigskip
{\bf Physical meaning of nonidentical evolutionary relation.}

The nonidentity of evolutionary relation means that
the balance conservation law equations are inconsistent. And this
indicates that the balance conservation laws are noncommutative. (If the balance
conservation laws be commutative, the equations would be consistent and the
evolutionary relation would be identical).

To what such noncommutativity of the balance conservation laws leads?

Nonidentical evolutionary relation obtained from the equations of
the balance conservation laws involves the functional that
specifies the material system state. However, since this relation
turns out to be not identical, from this relation one cannot get
the differential $d\psi $  that could point out to the equilibrium
state of material system. The absence of differential means that
the system state is nonequilibrium. That is, due to
noncommutativity of the balance conservation laws the material
system state turns out to be nonequilibrium under effect of
external actions. This points out to the fact that in material
system the internal force acts.

The action of internal force leads to a distortion of trajectories of material
system. The manifold made up by the trajectories (the accompanying
manifold) turns out to be a deforming manifold. The differential
form $\omega$,  as well as the forms $\omega^p$ defined on such manifold,
appear to be evolutionary forms. Commutators of these forms will contain
an additional term connected with the commutator of unclosed metric form of
manifold.

\bigskip
{\bf Selfvariation of nonidentical evolutionary relation.
(Selfvariation of nonequilibrium state of material system.)}

The availability of two terms in the commutator of the form $\omega^p $
and the nonidentity of evolutionary relation lead to that the relation
obtained from the balance conservation law equations turns out to be a selfvarying
relation.

Selfvariation of the nonidentical evolutionary relation points to the
fact that the nonequilibrium state of material system turns out
to be selfvarying. State of material system changes but remains nonequilibrium
during this process.

It is evident that this selfvariation proceeds under the action of internal force
whose quantity is described by the commutator of the unclosed evolutionary form
$\omega^p $. (If the commutator
be zero, the evolutionary relation would be identical, and this would
point to the equilibrium state, i.e. the absence of internal forces.)
Everything that gives a contribution into the commutator of the form
$\omega^p $ leads to emergency of internal force.

What is the result of such a process of selfvarying the nonequilibrium state
of material system?

\bigskip
{\bf Degenerate transformation. Emergency of closed exterior forms.
(Origination of physical structures.)}

The significance of the evolutionary relation selfvariation consists in
the fact that in such a process it can be realized conditions under
which the closed exterior form is obtained from the evolutionary form.

These are conditions of degenerate transformation.
Since the differential of evolutionary form, which is unclosed, is nonzero,
but the differential of closed exterior form equals zero, the transition from
evolutionary form to closed exterior form is possible only as a degenerate
transformation, namely, a transformation that does not conserve the differential.
And this transition is possible exclusively to inexact closed exterior form,
i.e. to the external form being closed on pseudostructure. The conditions of degenerate
transformation are those of vanishing the commutator of the metric form defining the
pseudostructure, in other words, the closure conditions for dual form.

As it has been already mentioned, the evolutionary differential form
$\omega^p$, involved into nonidentical relation (17) is an unclosed one.
The commutator of this form, and hence the differential, are nonzero.
That is,
$$d\omega^p\ne 0 \eqno(18)$$
If the conditions of degenerate transformation are realized, then from
the unclosed evolutionary form one can obtain a differential form closed
on pseudostructure. The differential of this form equals zero. That is,
it is realized the transition

$d\omega^p\ne 0 \to $ (degenerate transformation) $\to d_\pi \omega^p=0$,
$d_\pi{}^*\omega^p=0$

The relations obtained
$$d_\pi \omega^p=0,  d_\pi{}^*\omega^p=0 \eqno(19)$$
are the closure conditions for exterior inexact form. This means that
it is realized an exterior form closed on pseudostructure.

The realization of the closed (on pseudostructure) inexact exterior form
points to emergency of physical structure.

The conditions of degenerate transformation that lead to emergency of the
closed inexact exterior form are connected with any symmetries. Since these
conditions are conditions of vanishing the interior differential of the metric form,
i.e. vanishing the interior (rather than total) metric form commutator, the
conditions of degenerate transformation can be caused by symmetries of
coefficients of the metric form commutator (for example, these can be the
symmetrical connectednesses).

While describing material system the symmetries can be due to
degrees of freedom of the material system.
The translational degrees of freedom, internal degrees
of freedom of the system elements, and so on can be examples of
such degrees of freedom.

To the degenerate transformation it must
correspond a vanishing of some functional
expressions, such as Jacobians, determinants, the Poisson
brackets, residues and others. Vanishing these
functional expressions is the closure condition for dual form.
And it should be emphasize once more that {\it the degenerate
transformation is realized as a transition from the accompanying
noninertial coordinate system to the locally inertial system}.
The evolutionary form and nonidentical evolutionary relation
are defined in the noninertial frame of reference (deforming manifold). But
the closed exterior form obtained and the identical relation are obtained
with respect to the locally-inertial frame of reference (pseudostructure).

\bigskip
{\bf Obtaining identical relation from nonidentical one.
(Transition of the material system into a locally equilibrium
state.)}

On the pseudostructure $\pi$ evolutionary relation (17) converts into
the relation
$$
d_\pi\psi=\omega_\pi^p\eqno(20)
$$
which proves to be an identical relation. Indeed, since the form
$\omega_\pi^p$ is a closed one, on the pseudostructure this form turns
out to be a differential of some differential form. In other words,
this form can be written as $\omega_\pi^p=d_\pi\theta$. Relation (20)
is now written as
$$
d_\pi\psi=d_\pi\theta
$$
There are differentials in the left-hand and right-hand sides of
this relation. This means that the relation is an identical one.

From evolutionary relation (17) it is obtained the relation identical on
pseudostructure. In this case the evolutionary relation itself
remains to be nonidentical one. (At this point it should be
emphasized that the differential, which equals zero, is an interior one.
The evolutionary form commutator becomes zero only on the
pseudostructure. The total evolutionary form commutator is nonzero. That
is, under degenerate transformation the evolutionary form differential
vanishes only on pseudostructure. The total differential of the
evolutionary form is nonzero. The evolutionary form remains to be
unclosed.)

From relation (20) one can obtain the differential which specifies the
state of material system (and the state function), and this
corresponds to equilibrium state of the system.
But identical relation can be realized only on pseudostructure (which is
specified by the condition of degenerate transformation). This
means that a transition of material system to equilibrium state
proceeds only locally. In other words, it is realized a transition
of material system from nonequilibrium state to locally
equilibrium one. In this case the global state of material system
remains to be nonequilibrium.

The transition from nonidentical relation (17) obtained from
the balance conservation laws to identical
relation (20) means the following. Firstly, an emergency of the
closed (on pseudostructure) inexact exterior form (right-hand side
of relation (20)) points to an origination of the physical structure.
And, secondly, an existence of the state differential (left-hand side 
of relation (20))
points to a transition of the material system from nonequilibrium state
to the locally-equilibrium state.

Thus one can see that the transition of material system from
nonequilibrium state to locally-equilibrium state is accompanied
by originating differential-geometrical structures, which are
physical structures. The emergency of physical structures in the
evolutionary process reveals in material system as an emergency of
certain observable formations, which develop spontaneously.
Such formations and
their manifestations are fluctuations, turbulent pulsations,
waves, vortices, creating massless particles, and others. The
intensity of such formations is controlled by a quantity
accumulated by the evolutionary form commutator at the instant in
time of originating physical structures.

Physical structures that are generated by material systems made up
pseudometric manifolds on which physical fields are defined. 
Pseudo-Riemann and
pseudo-Euclidean manifolds are examples of such manifolds.

The availability of physical structures points out to the fulfilment
of conservation laws. These are conservation laws for physical fields.
The process of generating physical fields demonstrates a connection of
these conservation laws, which had been named as exact ones, with
the balance (differential) conservation laws for material media.
The closed inexact exterior forms that correspond to physical structures
and exact conservation laws are obtained from evolutionary forms which in turn
are obtained from the equations describing the balance conservation laws for
material media.

\bigskip
{\large\bf Characteristics and classification of physical structures.
(Characteristics and parameters of differential forms)}

Since the closed exterior form was obtained from the evolutionary form, 
it is evident that
characteristics of this structure has to be connected with those of
the evolutionary form and of the manifold on which this form is defined,
with the conditions of degenerate transformation and with the values of
commutators of the evolutionary form and the manifold metric form.

The conditions of degenerate transformation, i.e. the symmetries caused by degrees
of freedom of material system, determine the equation for pseudostructures.

The closed exterior forms corresponding to physical structures are
conservative quantities. These conservative quantities describe
certain charges.

The first term of the evolutionary form commutator determines the value
of a discrete change (the quantum),
which the quantity conserved on the pseudostructure undergoes during
transition from one pseudostructure to another. The second term of the
evolutionary form commutator specifies a characteristics that fixes the
character of the initial manifold deformation, which took place before
the physical structure emerged.  (Spin is an example of
such a characteristics).

Since the closed exterior forms corresponding to physical structures are
obtained from the evolutionary forms describing material systems, the
characteristics of physical structures depend on the properties of material
system generating these structures.

The closed exterior forms obtained correspond to the
state differential for material system. The differentials of entropy,
action, potential and others are examples of such differentials.

The connection of the physics structures with the
skew-symmetric differential forms allows to introduce a classification
of these structures in dependence on parameters that specify the
skew-symmetric differential forms and enter into nonidentical and
identical relation. To determine these parameters one has to consider
the problem of integration of the nonidentical evolutionary relation.

Under degenerate transformation from the nonidentical evolutionary
relation one obtains a relation being identical on pseudostructure.
Since the right-hand side of such a relation can be expressed in terms
of differential (as well as the left-hand side), one obtains a relation
that can be integrated, and as a result he obtains a relation with the
differential forms of less by one degree.

The relation obtained after integration proves to be nonidentical
as well.

By sequential integrating the nonidentical relation of degree $p$ (in
the case of realization of corresponding degenerate transformations
and forming the identical relation), one can get a closed (on the
pseudostructure) exterior forms of degree $k$, where $k$ ranges
from $p$ to $0$.

In this case one can see that after such integration the closed (on the
pseudostructure) exterior forms, which depend on two parameters, are
obtained. These parameters are the degree of evolutionary form $p$
in the evolutionary relation and the degree of created closed forms $k$.

In addition to these parameters, another parameter appears, namely, the
dimension of space. If the evolutionary relation generates the closed
forms of degrees $k=p$, $k=p-1$, \dots, $k=0$, to them there correspond
the pseudostructures of dimensions $(N-k)$, where $N$ is the dimension the
space formed.

The parameters of evolutionary and exterior forms that follow from
the evolutionary forms allow to introduce a classification
of physical structures that defines a type of physical structures
and, accordingly, of physical fields and interactions.

The type of physical structures generated by the evolutionary
relation depends on the degree of differential forms $p$ and $k$
and on the dimension of original inertial space $n$ (here $p$ is
the degree of the evolutionary form in the nonidentical relation
that is connected with the number of interacting balance
conservation laws, and $k$ is the  degree of closed form
generated by the nonidentical relation).

By introducing the classification by numbers $p$, $k$ and $n$ one can
understand the internal connection between various physical fields.
Since physical fields are the carriers of interactions, such
classification enables one to see a connection between
interactions.  It can be shown that the case $k=0$ corresponds to
strong interaction, $k=1$ corresponds to weak interaction,
$k=2$ corresponds to electromagnetic interaction, and $k=3$ corresponds
to gravitational interaction.

\bigskip
\centerline {\large\bf Conclusion} 
Invariant properties of closed exterior forms, which correspond to the 
conservation laws for physical fields, form the foundations of field 
theory. The field theory operators are built on the basis of 
transformations of closed exterior forms. These are gauge 
transformations. As it has been already noted, the field theory 
equations are connected with closed exterior forms. 

The closure of exterior forms that gives rise to the invariant 
properties is a result of conjugacy of the exterior form elements. 
The conditions of such a conjugacy are the availability of symmetries 
of exterior forms.

Evolutionary forms, which are obtained from the equations for material 
media, answer the question of how are realized the closed exterior 
forms that correspond to field theories.

This explains the process of originating physical fields and gives the 
answer to many questions of field theory.

Firstly, this shows that physical fields are generated by material media. 
And thus the causality of physical processes and phenomena is explained. 

Secondly, it becomes clear a connection of field theory with the equations 
of mathematical physics. The meaning of postulates, which the field theory 
is built on, is disclosed. The postulates, which are the conditions of 
conjugacy of various operators or objects, are the 
closure conditions of exterior forms. In field theories these conjugacy 
conditions (postulates) are imposed on the differential equations 
describing physical fields from the outside. But it turns out that these 
conditions are realized in the process described by the evolutionary 
forms that are obtained from the equations of mathematical physics for 
material media.

The connection between closed exterior forms corresponding to field 
theories and the evolutionary forms obtained from the equations for 
material media discloses a meaning of the field theory parameters. 
They relate to the number ($p$) of interacting balance conservation 
laws and to the degrees ($k$) of closed exterior forms realized. 
Hence it arises a possibility to classify physical 
fields and interactions according to the parameters $p$ and $k$. 

The connection of closed exterior forms with evolutionary forms 
discloses a relation between interior and exterior symmetries of field 
theories. Interior symmetries in field theories - symmetries of physical 
structures - are symmetries of closed inexact exterior forms. The 
exterior symmetries of the field theory equations are those for 
corresponding dual forms. They are caused by the degrees of freedom of 
material media generating physical fields.

Realization of symmetries of the exterior and dual forms described by 
the evolutionary forms leads to the fulfilment of conservation laws. 
This points out to emergency of conjugated objects, namely, physical 
structures. In material media, 
which generate physical structures, the emergency of physical 
structures manifests itself as an emergency of observable formations 
like fluctuations, pulsations, waves, vortices, massless particles, and 
so on.  

One of the problems in the theory of symmetry is a searching for 
symmetries of differential equations. A knowledge of symmetries enables 
one to get a solution of differential equations that corresponds to the 
conservation law and defines the differential-geometrical structures. 
The dependence of symmetries on any parameter, i.e. the dependence 
of the nondegenerate transformations on a given parameter allows to 
study the ``evolution" of the differential-geometrical structures 
in given parameter. But in this case the question of how these structures 
emerge is not posed (that is, the evolutionary process of originating 
these structures is not considered). As it has been shown in present 
paper, the answer to this question gives the theory of evolutionary 
forms.

It should be emphasized once more, that 
the necessity of using  evolutionary forms when describing physical 
processes is due to the fact that they make it possible to keep track 
of the conjugacy of the equations that describe physical processes. If 
these equations (or derivatives with respect to different variables) be 
not conjugated, then the solutions to corresponding equations prove to 
be noninvariant: they are functionals rather then functions. 
The realization of the conditions (while varying variables), 
under which the equations become conjugated ones (the realization of 
symmetries), leads to that the relevant solution becomes invariant.
Similar functional properties have the solutions to all differential 
equations. 

In the book the role of exterior and evolutionary differential forms 
for the field theory is emphasized. The exterior differential 
forms, whose properties correspond to the conservation laws, constitute 
the basis of the existing invariant field theories. They describe 
the physical structures, which constitute physical fields. However, the 
theory of exterior differential forms, being invariant one, cannot 
describe the processes of emergence of physical structures and formation 
of physical fields. This can be done only by the evolutionary theory. It 
is evident that the general field theory has to integrate the invariant 
and evolutionary mathematical foundations. It is an approach to such 
a general field theory that the theory of exterior and evolutionary 
differential forms can become.

1. Schutz B.~F., Geometrical Methods of Mathematical Physics. Cambrige 
University Press, Cambrige, 1982.

2. Bott R., Tu L.~W., Differential Forms in Algebraic Topology. 
Springer, NY, 1982.

3. Encyclopedia of Mathematics. -Moscow, Sov.~Encyc., 1979 (in Russian).

\end{document}